\documentstyle[a4p,12pt,amsbsy,epsf]{article}

\newcommand {\fittaubsRAW} {\mbox{$1.50^{+0.16}_{-0.15}$}}
\newcommand {\fittaubs}    {\mbox{$1.50^{+0.16}_{-0.15}$}} 
\newcommand {\tauBssys}    {\mbox{$0.04$}}

\newcommand {\fittaulbRAW} {\mbox{$1.26^{+0.23}_{-0.20}$}} 
\newcommand {\fittauLbval} {\mbox{$1.29$}} 
\newcommand {\fittauLbXibval} {\mbox{$1.27$}} 
\newcommand {\fittauLberr} {\mbox{$^{+0.24}_{-0.22}$}} 
\newcommand {\fittauLbXierr} {\mbox{$^{+0.23}_{-0.20}$}} 
 
\newcommand {\tauLbsys}    {\mbox{$0.06$}}
\newcommand {\XibLbsys}     {\mbox{$+0.02\pm0.02$}}
\newcommand{\fittauLbpKpi} {\mbox{$1.36^{+0.25}_{-0.23}$}} 
\newcommand{\fittauLblaml} {\mbox{$0.85^{+0.53}_{-0.37}$}} 
\newcommand {\tauLblamllsys}    {\mbox{$^{+0.11}_{-0.14}$}}
\newcommand {\nbsphysBG}     {\mbox{$27 \pm 11$}}
\newcommand {\nlbphysBG}     {\mbox{$16 \pm 7$}}
\newcommand {\ncanddsl}   {\mbox{199}}   
\newcommand {\ncanddslerr}{\mbox{26}}    
\newcommand {\ncandlcl}   {\mbox{145}}   
\newcommand {\ncandlclerr}{\mbox{24}}    

\newcommand {\ncandbs}     {\mbox{172}}  
\newcommand {\ncandbserr}  {\mbox{28}}   
\newcommand {\ncandlb}     {\mbox{129}}  
\newcommand {\ncandlberr}  {\mbox{25}}   

\newcommand {\ncandDLbs}  {\mbox{509}}   
\newcommand {\ncandDLlb}  {\mbox{632}}   
\newcommand {\nGPMH}      {\mbox{4.4}}   
\newcommand {\Zzero} {\mbox{$\rm{ Z}^0$}}

\newcommand {\bbar}      {{\mathrm{\overline{b}}}}
\newcommand {\lm}         {\ell^-}
\newcommand {\lp}         {\ell^+}
\newcommand {\p}          {\mathrm{p}}
\newcommand {\K}          {\mathrm{K}}
\newcommand {\D}          {\mathrm{D}}
\newcommand {\B}          {\mathrm{B}}
\newcommand {\Bb}         {\mathrm{\overline{B}}}
\newcommand {\X}          {\mathrm{X}}
 
\newcommand {\nubar}      {\bar{\nu}}
\newcommand {\Dsp}        {\mbox{$\mathrm{ D_s^+}$}}
\newcommand {\Dsm}        {\mbox{$\mathrm{ D_s^-}$}}
\newcommand {\Ds}         {\Dsm}
\newcommand {\ds}         {\mathrm{D_s}} 

\newcommand {\Lcp}        {\mathrm{\Lambda_c^+}}
\newcommand {\Lcm}        {\mathrm{\Lambda}_c^-}
\newcommand {\Lc}         {\Lcp}
\newcommand {\Xic}        {\mathrm{\Xi_c}}
\newcommand {\Xicb}       {\mathrm{\overline{\Xi}_c}}
\newcommand {\Bu}         {\mbox{$\mathrm{ B}^+$}}
\newcommand {\Bd}         {\mbox{$\mathrm{ B^0}$}}
\newcommand {\Bbar}      {\mathrm{\bar{B}}}
\newcommand {\Bs}         {\mathrm{B_s^0}}

\newcommand {\Lb}         {\mathrm{\Lambda_b^0}}
\newcommand {\Xib}        {\mathrm{\Xi_b}}
\newcommand {\Sigb}        {\mathrm{\Sigma_b}}
\newcommand {\phipi}      {\phi\pi^-}
\newcommand {\Kst}        {\K^{*0}}
\newcommand {\KstK}       {\Kst\K^-}
\newcommand {\phil}       {\phi\lm}
\newcommand {\phill}      {\phi\lm\lp}
\newcommand {\philnu}     {\phi\lm\nubar}
\newcommand {\philnux}    {\phi\lm\nubar\X}
\newcommand {\KK}         {\K^+\K^-}
\newcommand {\KKpi}       {\K^+\K^-\pi^-}
\newcommand {\KKpil}      {\KKpi\lp}

\newcommand {\KKll}       {\KK\lm\lp}
\newcommand {\KKp}        {\KKpi}

\newcommand {\Ko}         {\mathrm{K^0_S}}
\newcommand {\KKo}        {\K^-\Ko}
\newcommand {\KKol}       {\K^-\Ko\lp}
\newcommand {\Dsl}        {\Dsm\lp}
\newcommand {\Dslwr}      {\Dsm\lm}

\newcommand {\Dslnux}     {\Dsm\lp\nu\X}
\newcommand {\pKpi}       {\p\K^-\pi^+}
\newcommand {\pKp}        {\p\K\pi}
\newcommand {\pKpil}      {\pKpi\lm}
\newcommand {\pKpl}       {\pKpil}

\newcommand {\laml}       {\Lambda\lp}
\newcommand {\lamll}      {\Lambda\lp\lm}

\newcommand {\lamllbr}    {[\Lambda\lp]\lm}
\newcommand {\lamllwrbr}    {[\overline{\Lambda}\lp]\lm}

\newcommand {\lamlnux}    {\Lambda\lp\nu\X}
\newcommand {\ppill}      {\p\pi^-\lp\lm}
\newcommand {\Lcl}        {\Lcp\lm}
\newcommand {\Lclwr}      {\Lcp\lp}

\newcommand {\Lclnux}     {\Lcp\lm\nubar\X}
\newcommand {\Br}         {\mathrm{Br}}
\newcommand {\dEdx}       {\mbox{${\mathrm{d}}E/{\mathrm{d}}x$}}
\newcommand {\etal}       {{et al.}}
\newcommand {\ps}         {\mbox{$\mathrm{\,ps}$}}

\newcommand {\mic}        {\mbox{$\mathrm{\,\mu m}$}}
\newcommand {\tauBs}      {\mbox{$\mathrm{\tau(B_s^0)}$}}
\newcommand {\tauLb}      {\mbox{$\mathrm{\tau(\Lb)}$}}

\newcommand{\LB}           {{\cal L}_i^{\B}}
\newcommand{\LDl}          {{\cal L}_i^{\D\ell(\Lambda\ell)}}
\newcommand{\Lcombi}       {{\cal L}_i^{\rm{comb}}}
\newcommand{\fbgi}         {f^{\rm{comb}}_i}
\newcommand{\tb}           {\tau_{\B}}
\newcommand{\pb}           {p_{\B}}
\newcommand{\mb}           {m_{\B}}
\newcommand{\pdi}          {p_{\D}^i}
\newcommand{\mdi}          {m_{\D}^i}

\newcommand{\li}           {L^i}
\newcommand{\sigLi}        {\sigma_{L}^i}
\newcommand{\fp}           {f_{bg}^+}
\newcommand{\tbgp}         {\tau_{bg}^+}
\newcommand{\tbgn}         {\tau_{bg}^-}
\newcommand {\MeVcc}      {{\rm{MeV}}/c^2}
\newcommand {\GeV}        {{\rm{GeV}}}
\newcommand {\GeVc}       {{\rm{GeV}}/c}
\newcommand {\GeVcc}      {{\rm{GeV}}/c^2}
\newcommand {\ts}         {\thinspace}
\newcommand {\bsym}       {\boldsymbol}
\newcommand {\Caption}[1]  {\caption[]{\small\protect{\parbox[t]{13cm}{#1} }} }
\newcommand {\downto}
        {\mbox{ \begin{picture}(14,10)
                   \put(0,10){\line(0,-1){5.0}}
                   \put(2,5){\oval(4,4)[bl]}
                   \put(1,0){\makebox(0,0)[bl]{$\rightarrow$}}
                \end{picture} }}
\begin{document}
\pagenumbering{roman}
 
\begin{titlepage}
 
  \begin{center}
    {\large EUROPEAN LABORATORY FOR PARTICLE PHYSICS}
  \end{center} 
  \begin{flushright}
   CERN-PPE/97-159 \\
   16th December 1997
  \end{flushright}

  \begin{center}{\LARGE\bf
    Measurements of the $\bsym\Bs$ and  $\bsym\Lb$ Lifetimes \\}
  \end{center}
 
  \vspace{0.3in}
  \begin{center}
    {\Large\bf The OPAL Collaboration \large}
  \end{center}
 
  \vspace{0.1in}
  \begin{abstract}
    This paper presents updated measurements of the lifetimes of the
    $\Bs$ meson and the $\Lb$ baryon using \nGPMH~million hadronic
    $\mathrm{Z}^0$ decays recorded by the OPAL detector at LEP from
    1990 to 1995.  A sample of $\Bs$ decays is obtained using $\Dsl$
    combinations, where the $\Ds$ is fully reconstructed in the
    {\mbox{$\phipi$}}, {\mbox{$\KstK$}} and {\mbox{$\KKo$}} decay
    channels and partially reconstructed in the {\mbox{$\philnux$}}
    decay mode.  A sample of $\Lb$ decays is obtained using $\Lcl$ 
    combinations,
    where the $\Lc$ is fully reconstructed in its decay to a
    {\mbox{$\pKpi$}} final state and partially reconstructed in the
    {\mbox{$\lamlnux$}} decay channel.  From $\ncandbs \pm
    \ncandbserr$ $\Dsl$ combinations attributed to $\Bs$ decays, 
     the measured lifetime is \vspace{0.1in} 
      \[ \tauBs = \fittaubs \pm \tauBssys \ps, \]
    where the errors are statistical and systematic, respectively.
    From the $\ncandlb \pm \ncandlberr$ $\Lcl$ combinations
    attributed to $\Lb$ decays, the measured lifetime is
    \vspace{0.1in} {\[ \tauLb = \fittauLbval\fittauLberr \pm \tauLbsys \ps, \] }
  where the errors are statistical and systematic, respectively.

  \end{abstract}

  \medskip

  \vfill
  \begin{center}
    (Submitted to Physics Letters)
  \end{center}
 
\end{titlepage}

\newpage
\clearpage

\begin{center}{\Large        The OPAL Collaboration
}\end{center}
{\small
\begin{center}{
K.\thinspace Ackerstaff$^{  8}$,
G.\thinspace Alexander$^{ 23}$,
J.\thinspace Allison$^{ 16}$,
N.\thinspace Altekamp$^{  5}$,
K.J.\thinspace Anderson$^{  9}$,
S.\thinspace Anderson$^{ 12}$,
S.\thinspace Arcelli$^{  2}$,
S.\thinspace Asai$^{ 24}$,
S.F.\thinspace Ashby$^{  1}$,
D.\thinspace Axen$^{ 29}$
% ...
}\end{center}\bigskip
$^{  1}$School of Physics and Astronomy, University of Birmingham,
Birmingham B15 2TT, UK
\newline
$^{  2}$Dipartimento di Fisica dell' Universit\`a di Bologna and INFN,
I-40126 Bologna, Italy
\newline
% ...
$^{  5}$Cavendish Laboratory, Cambridge CB3 0HE, UK
\newline
% ...
$^{  8}$CERN, European Organisation for Particle Physics,
CH-1211 Geneva 23, Switzerland
\newline
$^{  9}$Enrico Fermi Institute and Department of Physics,
University of Chicago, Chicago IL 60637, USA
\newline
% ...
$^{ 12}$Indiana University, Department of Physics,
Swain Hall West 117, Bloomington IN 47405, USA
\newline
% ...
$^{ 16}$Department of Physics, Schuster Laboratory, The University,
Manchester M13 9PL, UK
\newline
% ...
$^{ 23}$Department of Physics and Astronomy, Tel Aviv University,
Tel Aviv 69978, Israel
\newline
$^{ 24}$International Centre for Elementary Particle Physics and
Department of Physics, University of Tokyo, Tokyo 113, and
Kobe University, Kobe 657, Japan
\newline
% ...
$^{ 29}$University of British Columbia, Department of Physics,
Vancouver BC V6T 1Z1, Canada
\newline
% ...
\newline

\newpage
\clearpage
 
\pagenumbering{arabic}
\section{Introduction}
\label{sec:intro}
The lifetimes of b~flavoured hadrons are related both to the strengths
of the b~quark couplings to c and u quarks, described by the CKM
matrix elements $V_{\rm cb}$ and $V_{\rm ub}$, respectively, and to
the dynamics of b~hadron decays.  The spectator model assumes that the
light quarks in b and c~hadrons do not affect the decay of the heavy
quark, and thus predicts the lifetimes of all b~hadrons to be equal.
For charm hadrons this prediction is inaccurate; non-spectator
effects, such as interference between different decay modes, result in
a $\mathrm{D^+}$ lifetime approximately 2.5 times that of the
$\mathrm{D^0}$ and more than twice that of the $\Ds$~\cite{PDG}.
Models which attempt to account for non-spectator effects predict that
the differences among b~hadron lifetimes are much smaller than
those in the charm system due to the larger mass of the
b~quark~\cite{HQET,BIGI,NEUBERT}.  These models predict a difference
in lifetime between the $\B^+$ and $\Bd$ meson of several percent, and
between the $\Bs$ and $\Bd$ meson of about $1\%$~\cite{BIGI,NEUBERT}.
OPAL~\cite{opalBslife,opalBslife-DsInclusive}, and other
collaborations~\cite{bslifetime-no-opal}, have published measurements
of the $\Bs$ lifetime which are in agreement with these models.

Non-spectator decays of B mesons proceeding via W-exchange are
Cabibbo-allowed but are expected to be suppressed, relative to
spectator decays, by an amount depending on the ratio of the
initial-state meson mass and the final-state quark masses
(helicity-suppression).  This suppression does not occur for baryon
decay, therefore b~baryon lifetimes are expected to be shorter than
b~meson lifetimes.  This expectation is consistent with existing
lifetime measurements~\cite{PDG} for which
$\tau(\Lb)/\tau(\Bd)=0.73\pm{0.06}$~\cite{PDG}.  It is also supported
by a recent theoretical prediction~\cite{BIGI} which yields
$\tau(\Lb)/\tau(\Bd)$ of about 0.9. Reference~\cite{NEUBERT}
finds this ratio to be $0.98+{\cal O}(1/{m_{\rm b}^3})$.  Measurements
of the average b~baryon lifetime have been published based both on
analyses of $\Lambda\lm$ and $\Lcl$ correlations by
OPAL~\cite{opal-Lc-Lblife,opal-Lamlepton-Lblife} and other
collaborations~\cite{lambda_b-lifetime-no-opal}.  In both analyses, the 
dominant contribution is expected to come from
$\Lb$ baryons, though both $\Lambda\lm$ and $\Lcl$ combinations can
arise from the decays of other b~baryons.
The composition of each sample depends on the b~baryon production
fractions, but the $\Lcl$ correlations provide a
purer sample of $\Lb$ baryons.

This paper presents updated measurements of the $\Bs$ meson and $\Lb$
baryon lifetimes using $\Dsl$ and $\Lcl$ combinations
reconstructed from the full OPAL hadronic data sample collected
on or near the $\Zzero$ resonance.  These results supersede the
previous OPAL measurements using $\Dsl$~\cite{opalBslife} and
$\Lcl$~\cite{opal-Lc-Lblife} combinations.  The decay
channels used for these lifetime measurements are:\footnote{In this
  paper, charge conjugate modes are always implied.} \vspace{-20pt}
\begin{tabbing}
  \hspace{2cm} \= \hspace{3.7cm} \= \hspace{2cm} \= \hspace{2cm} \= \hspace{3cm} \\
  \> $\Bs \to \Ds\, \lp \, \nu \,\X$ \> \>
  \> $\Lb \to \Lc\, \lm \, \bar{\nu} \,\X$ \\
  \> $\phantom{\Bs \to \hspace{5pt}} \downto \K^{*0}\, \K^-$,
  \> $\K^{*0}\rightarrow \K^+\pi^-$ \>
  \> $\phantom{\Lb \to \hspace{5pt}} \downto \pKpi$ \\ 
  \> $\phantom{\Bs \to \hspace{5pt}} \downto \phi\, \pi^-$,
  \> $\phi\rightarrow \K^+\K^-$  \>
  \> $\phantom{\Lb \to \hspace{5pt}} \downto \Lambda\lp\nu \X$,
     \ \ \ \ \ \ $\Lambda\rightarrow {\rm p}\pi^-$ \\ 
  \> $\phantom{\Bs \to \hspace{5pt}} \downto \Ko\, \K^-$,
  \> $\Ko\rightarrow \pi^+\pi^-$  \>
  \> \\
  \> $\phantom{\Bs \to \hspace{5pt}} \downto \phi\lm\bar{\nu} \X $,
  \>$\phi\rightarrow \K^+\K^-$ \>
  \>
\vspace{-5pt}
\end{tabbing}
where $\ell$ is an electron or a muon.  In each case, the proper decay
time of the b~hadron is determined on an event-by-event basis using
measured decay lengths and estimates of the b~hadron energy.

The following section provides a brief description of the OPAL
detector.  The remaining sections describe the selection of $\Dsl$
and $\Lcl$ candidates, the determination of the b~hadron decay
lengths, the estimation of the b~hadron boost, the lifetime fits, the
results, and the systematic errors.

\section{The OPAL Detector}
\label{sec:detector}
 
The OPAL detector is described in detail in
reference~\cite{opaldet-opalsi-opalsi2}. The central tracking system
is composed of a precision vertex drift chamber, a large volume jet
chamber surrounded by a set of chambers which measure the
$z$-coordinate\footnote{The right-handed coordinate system is defined
  such that the $z$-axis follows the electron beam direction and the
  $x$-$y$~plane is perpendicular to it with the $x$-axis lying
  approximately horizontally.  The polar angle~$\theta$~is defined
  relative to the $+z$-axis, and the azimuthal angle~$\phi$~is defined
  relative to the $+x$-axis.} and, for the majority of the data used
in this analysis, a high-precision silicon microvertex detector. These
detectors are located inside a solenoidal coil. The detectors outside
the solenoid consist of a time-of-flight scintillator array and a lead
glass electromagnetic calorimeter with a presampler, followed by a
hadron calorimeter consisting of the instrumented return yoke of the
magnet, and several layers of muon chambers.  Charged particles are
identified by their specific energy loss per unit length, $\dEdx$, in
the jet chamber.  Further information on the performance of the
tracking and $\dEdx$ measurements can be found in
reference~\cite{opaljet-hid}.

\section{Monte Carlo Simulation}
\label{sec:MCsim}

Monte Carlo simulation samples of inclusive hadronic \Zzero\
decays and of the specific decay modes of interest are used to check
the selection procedure and lifetime fit procedure. These samples were
produced using the JETSET~7.4 parton shower Monte Carlo
generator~\cite{jetset} with the fragmentation function of Peterson
\etal ~\cite{peterson} for heavy quarks, and then passed through the
full OPAL detector simulation package~\cite{opalmc}. A special sample
of simulated data was generated using a modified JETSET decay routine
for b~baryons~\cite{jetpol}, where it is assumed that the polarization
of the b quark is carried by the b~baryon. An additional form 
factor~\cite{formf} describing the energy transfer from the b to c
flavoured baryon was used in the generation of the polarized sample.

\section{Candidate Selection}
\label{sec:selection}
 
This analysis uses data collected during the 1990--1995 LEP running
periods at centre-of-mass energies within $\pm3\,\GeV$ of the
$\mathrm{Z}^0$ resonance.  After the standard hadronic event
selection~\cite{opalmh} and detector performance requirements, a
sample of \nGPMH~million events is selected.  Jets are defined using
charged tracks and electromagnetic clusters not associated with a
charged track.  These are combined into jets using the scaled
invariant mass algorithm with the E0 recombination
scheme~\cite{bib-JADE} using $y_{\mathrm {cut}} = 0.04$.

Only charged tracks that are well-measured in the $x$-$y$ plane are
considered in this analysis.  Well measured tracks are defined
according to standard track quality cuts~\cite{opal-track-sel}.
Within a single jet, not all combinations of accepted tracks with the
appropriate charge combination are considered in the $\Ds$ and
$\Lc$ searches.  Instead, to reduce the combinatorial background, the
$\dEdx$ probability, $w_i$, that the observed $\dEdx$ is consistent
with the assumed particle hypothesis, $i$,
is required to be greater than $1\%$. 
More restrictive requirements are imposed on a
channel-by-channel basis, as described in the following subsections.

Leptons are identified as follows.  Electron candidates with a
momentum of at least $2\,\GeVc$ are identified using an artificial
neural network based on twelve measured quantities from the
electromagnetic calorimeter and the central tracking
detector~\cite{BdmixNNeID}.  Between 1 and 2 $\GeVc$, electron
candidates are required to have a $\dEdx$ probability of larger than
$1\%$ for the electron hypothesis and less than $1\%$ for the proton
hypothesis, because it is in this region that the electron and proton
$\dEdx$ bands cross.  Electron candidates identified as arising from
photon conversions are rejected~\cite{gambblep}.  Muons are identified
by associating central detector tracks with track segments in the muon
detectors and requiring a position match in two orthogonal
coordinates~\cite{gambblep}.
 
\subsection{Selection of $\Ds$ candidates}
\label{sec:Ds}
 
The $\Ds$ candidates are reconstructed in four modes:
\begin{enumerate}
\item $\Ds\to\KstK$ in which the $\Kst$ decays into a
$\K^+\pi^-$.  

\item $\Ds\to\phipi$ where the $\phi$ subsequently decays into $\KK$. 

\item $\Ds\to\KKo$ where the $\Ko$ decays into $\pi^+ \pi^-$.

\item The $\Ds$ is partially reconstructed in
      $\Ds \to \philnux$ where the 
      $\phi$ decays into $\K^+\K^-$.

\end{enumerate}

In all modes except $\philnux$, charged kaon candidates are required to have a
momentum greater than $2\,\GeVc$.  Also, because of the potential for
misidentifying a pion as a kaon, if the observed energy loss of a
kaon candidate is greater than the mean expected for a kaon, it must
satisfy $w_{\K} > 5\%$.


In both the $\Ds\to\KstK$ and the $\Ds\to\phipi$ channels,
if the observed energy losses of both kaon candidates are greater than
the mean expected for a kaon, the product of the two $\dEdx$
probabilities must satisfy $w_{\K 1}\cdot w_{\K 2}> 0.02$.  The
momentum of the $\KKpi$ combination is required to be greater than 
$9\,\GeVc$ for both channels.

For the $\KstK$ mode, the invariant mass of the $\K^+\pi^-$
combination is required to satisfy $0.845 < m_{\K\pi} <
0.945\,\GeVcc$.  To reduce the possibility of mistaking a
$\D^-\to\Kst\pi^-$ decay for the desired signal, the measured
$\dEdx$ of the $\K^-$ candidate must be at least one standard
deviation below the mean $\dEdx$ that is expected for a pion.

In the $\phipi$ mode, the observed $\phi$ width is dominated by
detector resolution and the $\KK$ mass is required to satisfy
$1.005 < m_{\K\K} < 1.035\,\GeVcc$. The momentum of the $\phi$
candidate is required to be greater than 4.0 $\GeVc$.

Differences between the angular distributions of $\Ds$ decays and
those of random track combinations are used to suppress further the
combinatorial background. The $\Ds$ is a spin-0 meson and the final
states of both decay modes consist of a spin-1 ($\phi$ or $\K^{*0}$)
meson and a spin-0 ($\pi^-$ or $\K^-$) meson.  The $\Ds$ signal is
expected to be uniform in $\cos\theta_p$, where ${\theta_{p}}$ is the
angle in the rest frame of the $\Ds$ between the spin-0 meson
direction and the $\Ds$ direction in the laboratory frame.  However, the
$\cos\theta_p$ distribution of random combinations peaks in the
forward and backward directions.  It is therefore required that
$|\cos\theta_p| < 0.8$ (0.9) for the $\KstK$ ($\phipi$) mode.  The
distribution of $\cos\theta_v$, where $\theta_v$ is the angle in the
rest frame of the spin-1 meson between the direction of the final
state kaon from the decay of the spin-1 meson and the $\Ds$ direction,
is proportional to $\cos^2\theta_v$ for $\Ds$ decays.  The
$\cos\theta_v$ distribution of the random $\KKp$ combinations in the
data is, however, approximately flat.  Therefore it is required that
$|\cos\theta_v| > 0.4$.


For reconstruction of the $\Ds\to\KKo$ channel, accepted charged kaons
are combined with $\Ko$ mesons reconstructed in their decays to
$\pi^+\pi^-$ as described in~\cite{kssel}. The $\pi^+\pi^-$ mass is
required to satisfy $0.475 < m_{\pi\pi}< 0.525\,\GeVcc$.  The
background of $\Ko$ particles from fragmentation is reduced by
requiring the $\Ko$ momentum to be greater than $3\,\GeVc$. 
To reject background from $\Lambda$ decays where the
proton is misidentified as a pion, $\Ko$ candidates are rejected if
$1.10 < m_{\p\pi} < 1.13\,\GeVcc$, where $m_{\p\pi}$ is the invariant
mass of the two tracks when the highest momentum track is assigned the
proton mass. To improve further the mass resolution of the $\Ds$, a
fit using kinematic and geometrical constraints is performed.  
The mass of the two tracks forming the $\Ko$ is constrained to the
known $\Ko$ mass~\cite{PDG}.
Further constraints are applied to the $\Ds$ and
$\Ko$, in which the directions of the vectors between their production and
decay points are constrained to be the same as the reconstructed
momentum vectors.

For the $\Ds\to\philnu\X$ selection, the purity of the kaons from the
$\phi$ decay is enhanced by applying additional pion rejection.  This
requires that the $\dEdx$ probability for the pion hypothesis be less
than $40\%$ for tracks whose momenta are such that the mean $\dEdx$ for
pions is higher than that of kaons, and less than $1\%$ for low
momentum tracks where the mean $\dEdx$ for kaons is higher than that
of pions.  Additionally, the product of the $\dEdx$ probabilities for
a pion hypothesis for the two kaon candidates must be less than 0.018.
The $\phi$ candidate momentum is required to be greater than $4.6\,\GeVc$.
The lepton candidate is required to have a momentum greater than
$1\,\GeVc$ for electrons and $2\,\GeVc$ for muons.  The invariant mass
of the $\phil$ combination must be less than $1.9\,\GeVcc$.

\subsection{Selection of $\Lc$ candidates}
\label{lcsel}
The $\Lc$ candidates are reconstructed in two modes; $\Lc\to\p\K^-\pi^+$
and $\Lc\to\Lambda\lp\nu \mathrm{X}$.
In the $\p\K^-\pi^+$ selection, the proton, kaon and pion momenta are
required to be greater than 3, 2 and 1 $\GeVc$, respectively. If the
observed energy loss of the proton or kaon candidate is greater than
the mean $\dEdx$ expected for that particle, the corresponding $\dEdx$
probability is required to be greater than 3\%.  The $\dEdx$
probability of the pion hypothesis for the proton candidate is
required to be less than 1\%, which substantially reduces the
combinatorial background.  The $\p\K^-\pi^+$ combination must have a
momentum greater than $9\,\GeVc$.  To reduce the possibility of mistaking the
decay $\Dsp\to\phi(\K^+\K^-)\pi^+$ for a $\Lc$ decay by
misidentifying the $\K^+$ as a proton, candidates are rejected if the
invariant mass of the $\p\K^-$ combination, when the proton candidate
is assigned the kaon mass, is within $10\,\MeVcc$ of the
nominal $\phi$ mass~\cite{PDG}.

In the $\Lc\to\lamlnux$ selection, $\Lambda$ candidates are identified
via the decay $\Lambda\to\p\pi^-$. The selection procedure is
similar to the one used in reference~\cite{opal-Lamlepton-Lblife}.
The track with the larger momentum is assumed to be the proton and its
momentum is required to be larger than $3.0$ $\GeVc$. The other track
is required to have a momentum larger than $0.8\,\GeVc$. The
selection criteria for the lepton are the same as described in the
previous section for the selection of $\Ds$ semileptonic decays. The
invariant mass of the $\laml$ combination is required to be
less than $2.2\,\GeVcc$. 

\subsection{$\bsym\Ds \bsym\lp$ and $\bsym\Lc \bsym\lm$ selection 
and decay length determination}
\label{seldl}
 
Once a combination of tracks that satisfies the $\Ds$ or $\Lc$
candidate selection is found, a search is performed to find a lepton
from b hadron decay 
of opposite charge in the same jet. Lepton candidates are identified
as described in the introduction to section~\ref{sec:selection}.  Both the 
electron and muon
candidates are required to have a momentum greater than $2\,\GeVc$
except in the $\Ds\to\KKo$ mode where a higher momentum cut of
$5\,\GeVc$ is used to improve the signal to background ratio.
It is also required that the lepton candidate track be measured
precisely by either the silicon microvertex detector or the vertex drift chamber.
 
To further suppress combinatorial background, requirements are made on
the invariant mass and momentum of the $\Dsl$ and $\Lcl$ candidate
combinations.  $\KKpil$ combinations are required to have a mass
between $3.2$ and $5.5\,\GeVcc$ and momentum larger than $17\,
\GeVc$.  $\KKol$ combinations are accepted if they have an invariant
mass between $3.4$ and $5.5\,\GeVcc$ and a momentum greater than 17
$\GeVc$.  The $\phill$ mass is required to be less than $4.8\,\GeVcc$
and its momentum greater than $12\,\GeVc$.  Also, for the $\phill$
mode, the invariant mass of the $\phi$ and the $\lp$ must be greater
than $2.1\,\GeVcc$.  
In conjunction with the invariant mass cut on the
$\phi\lm$ pair, this unambiguously separates the leptons from the
$\Bs$ and $\Ds$ decays.  
The $\pKpil$ combination must have a mass between $3.5$ and
$5.5\,\GeVcc$ and momentum greater than $17\,\GeVc$.  In the $\lamll$
channel, combinations are accepted if the $\lamll$ invariant mass is
greater than $2.5\, \GeVcc$ and less than $5\, \GeVcc$ and the
invariant mass of the $\Lambda$ and $\lm$ is greater than
$2.2\,\GeVcc$. 
Furthermore, the cosine of the opening
angle between the lepton and the $\Ds$ or $\Lc$ candidate must be
greater than 0.4.
 
Three vertices --- the beam spot, the $\Bs$ ($\Lb$) decay vertex and
the $\Ds$ ($\Lc$) decay vertex --- are reconstructed in the $x$-$y$
plane.  The beam spot is measured using charged tracks with a
technique that follows any significant shifts in the beam spot
position during a LEP fill~\cite{taulife}.  The intrinsic width of the
beam spot in the $y$ direction is taken to be $8\mic$.
The width in the $x$ direction is measured
directly and found to vary between $100\mic$ and $160\mic$.
 
The $\Ds$ ($\Lc$) vertex is fitted in the $r$-$\phi$ plane using all
the candidate tracks. The $\Bs$ ($\Lb$) decay vertex is formed by
extrapolating the candidate $\Ds$ ($\Lc$) momentum vector from its
decay vertex to the intersection with the lepton track.  The
$\Ds$ ($\Lc$) decay length is the distance between these two decay
vertices.  The $\Bs$ ($\Lb$) decay length is found by a fit between the
beam spot and the reconstructed $\Bs$ ($\Lb$) decay vertex using the
direction of the candidate $\Dsl$ ($\Lcl$) momentum vector as a
constraint.  The two-dimensional projections of the $\Bs$ ($\Lb$) and
$\Ds$ ($\Lc$) decay lengths are converted into three dimensions using
the polar angles that are reconstructed from the momenta of the
$\Dsl$ ($\Lcl$) and $\Ds$ ($\Lc$).
Typical decay lengths for the $\Dsl$ ($\Lcl$) vertex are about 0.3\,cm
and the corresponding decay length errors range from about 0.03\,cm
for the $\KKpi$, $\philnu$ and $\pKpi$ modes, to about twice this level
for the modes which include a $\Lambda$ or $\Ko$.
 
Additional criteria are used to select $\Dsl$ and $\Lcl$ candidates
suitable for precise decay length measurements.  In channels in which
a charm state is fully reconstructed the $\chi^2$ of the charm vertex
fit is required to be less than 10 (for 1 degree of freedom).
Finally, the decay length error of the reconstructed $\Bs$ ($\Lb$)
candidate must be less than $0.2\,\mathrm{cm}$.
 
\subsection{Results of $\bsym\Ds \bsym\lp$ and $\bsym\Lc \bsym\lm$ 
selections}
 
The invariant mass distributions obtained in each of the $\Ds$ decay
modes are shown in figure~\ref{fig:Dsmass}.  The equivalent
distributions for the reconstructed $\Lc$ decay modes are shown in
figure~\ref{fig:lcmass}.  
\begin{figure}[ptb]
  \begin{center}
  \begin{minipage}{0.9\textwidth}
     \epsfxsize=\textwidth
      \epsffile{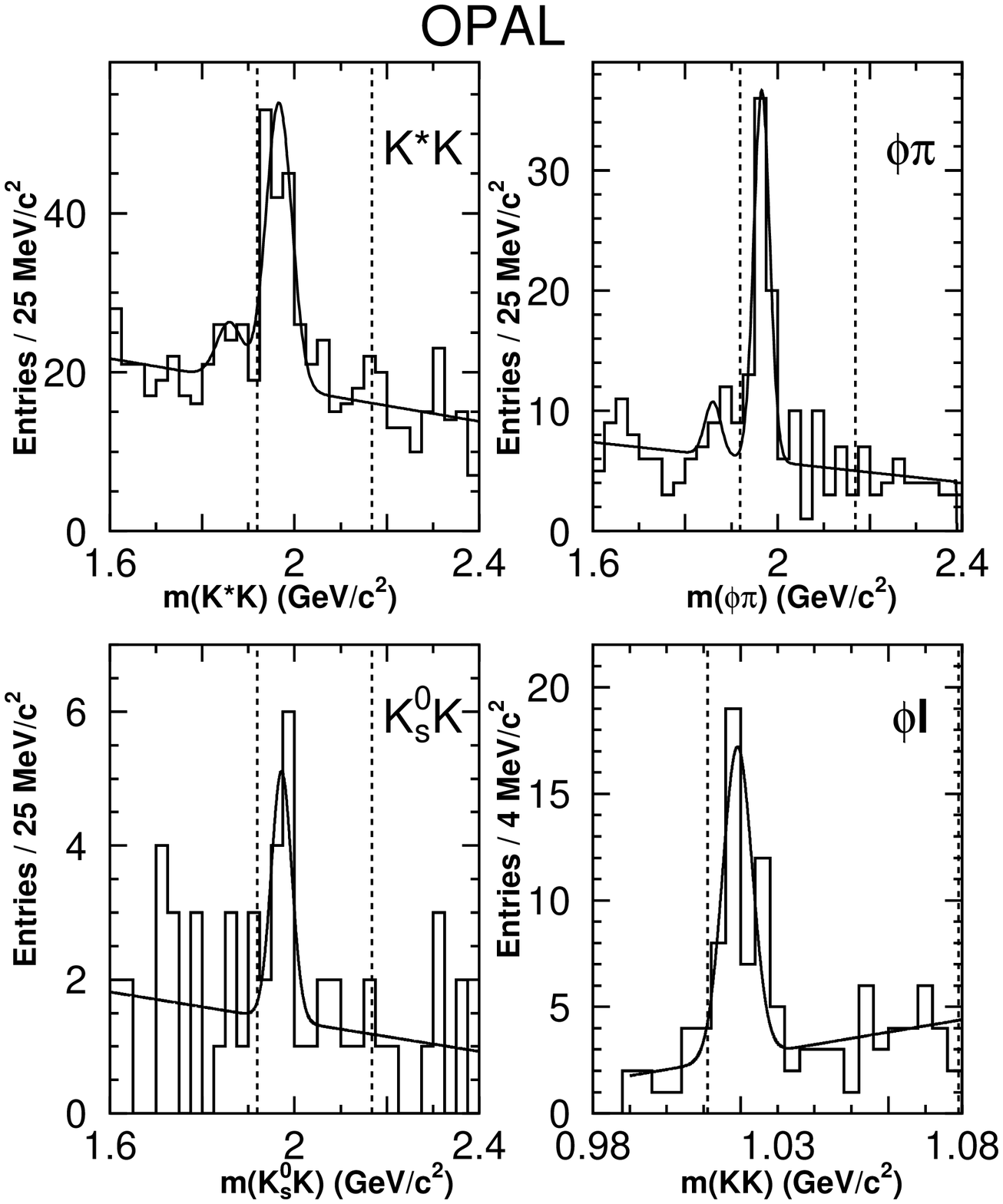}
  \end{minipage}
    \vspace{-0.2in} 
    \Caption{Invariant mass distributions from the 
      four $\Dsl$ reconstruction channels.  In each plot,
      the result of the fit described in the text is overlaid as 
      a solid line. The mass ranges used in the decay length fit are
      shown by the vertical dotted lines.  }
     \label{fig:Dsmass}
  \end{center}
\end{figure}
\begin{figure}[ptb]
  \begin{center}
  \begin{minipage}{0.9\textwidth}
     \epsfxsize=\textwidth
      \epsffile{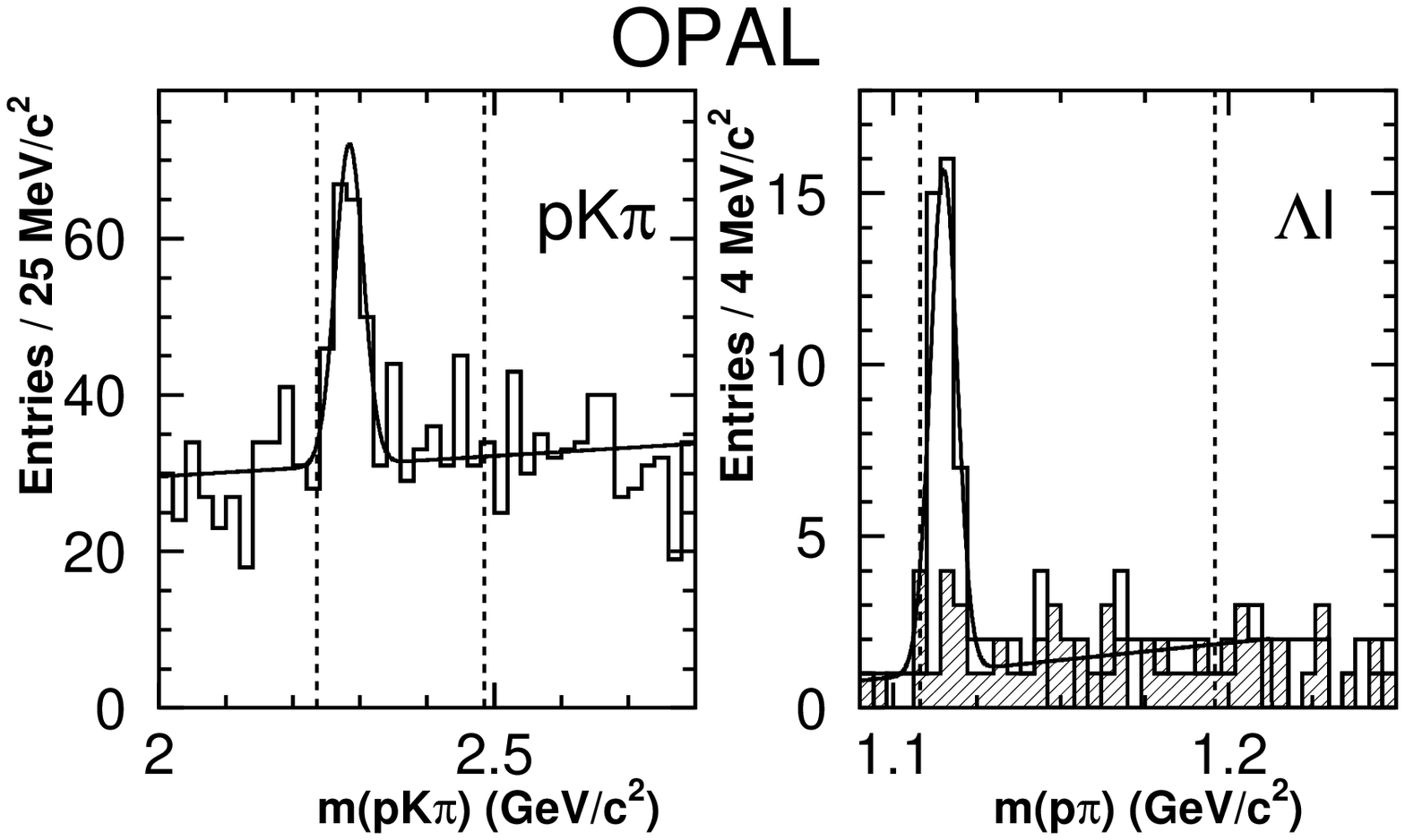}
  \end{minipage}
    \vspace{-0.2in} 
    \Caption{Invariant mass distributions
      from the two $\Lcl$ reconstruction channels.  In each plot,
      the result of the fit described in the text is overlaid as
      a solid line. Also indicated are the mass ranges used in the decay
      length fit. The hatched histogram for the $\lamllbr$ channel
      represents the wrong-sign-$\Lambda$ combinations: $\lamllwrbr$.}
     \label{fig:lcmass}
  \end{center}
\end{figure}
In each case, the fit result overlaid on the histogram is obtained from an 
unbinned maximum likelihood fit to the invariant mass distribution. 
The results of these fits are 
summarised in table~\ref{tab-signal}.

\begin{table}[bth]
  \centering
  \begin{tabular}{|l|c|c|c|c|r|}
    \hline
      Decay         &  Signal     & Comb. & Width & Fit Range& Cands. \\
      channel       &  candidates & fraction & (MeV) & (MeV)    & for fit \\
    \hline \hline
   $\K^*\K$   & 101$\pm$21 &  0.46$\pm0.06$ & 28$\pm$5 & $-$50 - 200 & 280\\
   $\phi\pi$  & 53$\pm$11  &  0.24$\pm$0.05 & 17$\pm$4 & $-$50 - 200 & 114\\
   $\KKo$     & 8$\pm$5    &  0.39$\pm$0.19 & 22$\pm$14& $-$50 - 200 &  21 \\
   $\phi\ell$ & 37$\pm$10  &  0.23$\pm$0.07 & 4 (fixed)&  $-$8 - 60  &  94\\
    \hline
   $\Dsl$ total  & $\ncanddsl\pm\ncanddslerr$
                   &                &          &          & \ncandDLbs  \\
    \hline \hline
   $\pKp$        & 108$\pm$22 &  0.56$\pm$0.06 & 21$\pm$5 & $-$50 - 200 & 522\\
   $\rm{\Lambda\ell}$ & 37$\pm$9&0.11$\pm$0.05&$3.4\pm0.5$& $-$8 - 80&69(41) \\
    \hline
     $\Lcl$ total  &  $\ncandlcl\pm\ncandlclerr$ 
                                &                &          &    &\ncandDLlb\\
    \hline
  \end{tabular}
  \Caption{
    Results of the mass fits to all the signal
    channels. The second column shows the
    number of candidates in the signal peak. The estimated fraction of
    combinatorial background is given in the third.  The fourth column
    gives the signal peak width found by the fit or the constraint
    that was used in the fit to the invariant mass distribution. The mass range
    used in the lifetime fit is given in the fifth column and the last
    column gives the number of candidates in this mass range.  For the
    $\lamllbr$ channel the number of wrong-sign-$\Lambda$
    candidates, $\lamllwrbr$, is included in brackets.}
    \label{tab-signal}
    \vspace{0.5cm}
\end{table}

Each mass fit uses a Gaussian function to describe the signals and a
linear parametrization of the combinatorial background. In the $\KKp$
distributions, a second Gaussian is used to parametrize contributions
from the Cabibbo suppressed decay $\D^-\to\KKpi$.  The mean of this
Gaussian is fixed to the nominal $\D^-$ mass,
$1869.3\,\MeVcc$~\cite{PDG}, and the width is constrained to be the
same as that of the $\Ds$ peak.  By integrating the tail of the peak
due to the $\D^-\to\KKpi$ decays in this $\Ds$ signal region, the
contamination from this source is found to be negligible.  No
significant peaks are observed in the mass distributions for
wrong-sign $\Dslwr$ ($\Lclwr$) combinations in the fully reconstructed
decay channels $\KstK$, $\phipi$, $\KKo$ and $\pKpi$.  The
$[\Lambda\lp]\lm$ combinations\footnote{ The bracketed particles are
  those that are assigned to be the decay products of a $\Lc$. The invariant
mass requirements on $\Lambda\ell$ combinations described in sections 4.2
and 4.3, result in a unique assignment of the two leptons.}
include a contribution from $\Lambda$ baryons from fragmentation that
can be estimated from the wrong-sign-$\Lambda$ distribution,
$\lamllwrbr$.  Studies using simulated data show that the
wrong-sign-$\Lambda$ distribution provides a good representation of
these $\Lambda$ baryons from fragmentation.  The $\lamllwrbr$
candidates are shown in the plot as a shaded histogram.

For each channel, the fitted mass is consistent with the nominal $\Ds$
($\Lc$) mass~\cite{PDG} and the fitted width is consistent with the
expected detector resolution. In total, $\ncanddsl\pm\ncanddslerr$
$\Dsl$ candidates and $\ncandlcl\pm\ncandlclerr$ $\Lcl$ candidates are
observed.  The $\Dsl$ ($\Lcl$) combinations used in the $\Bs$ ($\Lb$)
lifetime fits are selected from regions around the identified
invariant mass peaks, including a sufficient number of candidates away
from the mass peaks to allow an estimate of the lifetime characteristics
of the combinatorial background. There are $\ncandDLbs$ for the $\Bs$
lifetime fit and $\ncandDLlb$ for the $\Lb$ lifetime fit.

\subsection{Backgrounds to the
  $\bsym\Bs \to \bsym\Ds \bsym\lp$ and 
  $\bsym\Lb \to \bsym\Lc \bsym\lm$ signal}
\label{sec:phys-bg}
 
Potential sources of backgrounds to the $\Bs$ ($\Lb$) signal
considered here include decays of other b~hadrons that can yield a
$\Dsl$ ($\Lcl$) final state or other final states that are
misidentified as a $\Ds$ ($\Lc$) hadron.  Other sources are $\Ds$
($\Lc$) hadrons combined with a hadron that has been misidentified
as a lepton, and random associations of $\Ds$ ($\Lc$) hadrons with
genuine leptons. Finally, there is purely combinatorial background.
The various physics backgrounds, and the calculation of their
contributions relative to that of the signal, are discussed below.
 
\subsubsection {Physics backgrounds to $\bsym\Bs \to \bsym\Ds \bsym\lp$}

The signal event samples include properly reconstructed
$\Dsl$ combinations that do not arise from $\Bs$ decay.  Two decay
modes of $\Bd$ and $\Bu$ mesons are considered: 
\begin{itemize}
  \item[(a)] $\overline{\B} \to \Ds \D \X$, 
             \mbox{$\D\to\lp\nu\X$} 
             (where $\D$ is any non-strange charm meson), and
  \item[(b)] \mbox{$\B\to \Ds \K \lp\nu\X$}, 
             where $\K$ is any type of kaon.
\end{itemize}
For the signal production and decay sequence, the production rate
times branching fraction $f(\bbar\to\Bs)\,\cdot$
\mbox{$\Br(\Bs\to\Dslnux) =$} \mbox{$0.85 \pm 0.23\%$} is
used~\cite{PDG}.  Monte Carlo simulations are used to determine the
selection efficiencies for background modes relative to that of the
signal mode.  
 
For the background, the probability for a bottom quark to form either
a $\Bu$ or $\Bd$ meson is $0.378\pm0.022$~\cite{PDG} each.  For the
$\B\to\Ds\K\lp\nu\X$ mode, the measured branching ratio is
less than $0.009$ at the 90\% confidence level~\cite{PDG}.  
Half of this limit
is used as a central value in estimating the contribution of this
channel, and the range from zero to 0.009 is taken as the uncertainty.
For the other
background mode, it is noted that $\Br(\B\to\ds\D) =
0.049\pm0.011$~\cite{PDG}, which is then corrected using Monte Carlo
simulation to include the additional contribution from $\B\to\ds\D\X$
decays.  The possibility that all the $\B\to\ds\X$
modes include an additional charm meson is considered as a systematic
error.  The effect on the reconstruction efficiency of 
$\overline{\D}$ mesons arising from orbitally excited $\D$ mesons 
is also taken into account. 
The total contribution from these two sources
of backgrounds is estimated to be $11\pm4\%$.


\subsubsection {Physics backgrounds to $\bsym\Lb \to \bsym\Lc \bsym\lm$}

The events in the $\Lc$ peak may include $\Lcl$ combinations that do
not arise from $\Lb$ decay.  The decay modes considered are
$\B_{\mathrm{u,d}}\to\Lc\Xicb\X$, $\Xicb\to\X\lm\nubar$ and
$\Bb_{\mathrm{u,d,s}}\to\Lc\X\lm{\nubar}$.  An estimate of the $\Lb$
signal contamination from other b baryons is also given.

As before, the reconstruction efficiencies for these background modes are
calculated relative to the signal mode from simulated event samples.  For 
the production rate times branching fraction of the signal modes, 
$f(\bbar\to\Lb)\cdot\Br(\Lb\to\Lclnux) = 1.35\pm0.26\%$~\cite{PDG} is used.  

To estimate the background from the internal-W decay $\B\to\Lc\Xicb\X$,
$\Xicb\to\X\lm\bar{\nu}$, the measured inclusive branching ratio
$\Br(\B\to\mathrm{\mbox{charmed-baryon}}~\X) = 6.4\pm1.1\%$~\cite{PDG}
is combined with the measurement~\cite{cleo-BtoLamc}
$\Br(\B\to\Lc\X)/\Br(\B\to\Lcm\X)=0.19\pm0.13\pm0.04$,
where $\mathrm{B}$ refers only to $\mathrm{B}$ mesons containing a
$\bar{\rm{b}}$ quark. 
It is also assumed that when a $\Lc$ is
produced, a $\Xicb$ is always produced.  
The average semileptonic branching ratio of the charged and neutral
$\Xic$ (assuming they are produced at equal rates) is estimated to be
$25\pm 10\%$.  This was obtained from the semileptonic branching
ratio of the $\Lc$, using the theoretical prediction of
\cite{Xic-theory} and the  measured lifetimes of these
baryons~\cite{PDG}.
After accounting for the relative efficiency, this mode
comprises $2.0\pm1.5\%$ of the signal.

The contribution from the external-W decay, $\Bb\to\Lc\X\lm\nu $, was
estimated using the 90\% confidence level limit
$\Br(\Bbar_{\rm u,d}\to\p\lm\nubar\X)<0.16\%$~\cite{argus-BtopeX}.  It is
conservatively assumed that this decay always proceeds through a $\Lc$
and that a $\Lc$ is equally likely to decay to a proton or a
neutron.  Thus, $\Br(\B\to\Lc\lm\nubar\X) < 0.32\%$ at the 90\%
confidence level.  The analogous decays of $\Bs$ mesons are also taken
into account by assuming that their branching fraction to
$\Lc\lm\nubar\X$ is the same as for $\B_{\mathrm{u,d}}$ and using
the hadronization fractions from~\cite{PDG}. Accounting for the
relative detection efficiency yields an upper limit for this mode
corresponding to 5.7\% of the signal.  Half of this is taken as the
central value for this background and the entire range from zero to
5.7\% is considered in estimating the systematic error.

Potential contamination of the $\Lcl$ signal by decays of b~baryons
other than $\Lb$ is also investigated.  The principle sources of this
contamination are from decays of $\Xib$ and $\Sigb$.  There is some
evidence for the $\Xib$~\cite{Xi_b}, which is expected to decay
weakly~\cite{b-baryon-masses}.  Theoretical
predictions~\cite{b-baryon-masses} for the $\Sigb$ mass suggest that
it is large enough to allow strong decay to a $\Lb$.  Accepting this,
any non-$\Lb$ in the signal comes from $\Xib$ decays.

Semileptonic decays of $\Xib$ baryons to excited charm-strange baryons
that decay subsequently to $\Lc$, or non-resonant decays such as
$\Xib\to\Lc\X\lm\nubar$ can contribute to the $\Lcl$ sample.  The
level of these decays is estimated using the B meson system as a
guide.  The branching ratio for non-strange B decays to $\X\lp\nu$,
where X is not simply a D or $\D^*$ meson, is found to be
$3.7\pm1.3\%$.  This value is obtained by subtracting
$\Br(\Bd\to\D^-\lp\nu)$ and $\Br(\Bd\to\D^{*-}\lp\nu)$ from
$\Br(\Bd\to\X\lp\nu)$ using the values from reference~\cite{PDG}.  The
same rate is assumed for the analogous $\Xib$ decays mentioned above
and the conservative assumption is made that in these decays a $\Lc$
is always produced.  It is further assumed that $20\%$ of the weakly
decaying b baryons are
$\Xib$, based on the relative rates of production of the 
corresponding 
light-flavoured baryons in $\rm Z^0$ decays~\cite{PDG}.
Using the above branching ratio estimates and
reconstruction efficiencies for the signal mode and the modes under
consideration here, the contribution to the $\Lcl$ signal from the
decays of the $\Xib$ is estimated to be about 1\% and is, therefore,
neglected.

The $\lamll$ sample can have additional contributions from the decay
of $\Xib\to\Xic\lm\nubar\X$ followed by $\Xic\to\Xi\lp\nu\X$ with
$\Xi\to\Lambda\pi$.  
Assumptions about ${\rm
  Br(\Lc\to\Lambda\X\ell^+\nu)/Br(\Xic\to\Xi\X\ell^+\nu)}$ and $\rm
Br(\Xic\to\Xi\X)/Br(\Lc\to\Lambda\X)$ are required to estimate this
background if measurements of
$\Lambda\ell$~\cite{opal-Lamlepton-Lblife,lambda_b-lifetime-no-opal}
and $\Xi\ell$~\cite{Xi_b} production are to be used.
In the case of the former ratio, this cannot be trivially related to
the ratio of lifetimes because of Pauli Interference effects between
the two strange quarks that result from the decay of a $\Xic$, but
which are not present in the analogous $\Lc$ decay~\cite{Xic-theory}.
The latter ratio is even harder to predict theoretically, even to the
extent of whether one would expect it to be greater than or less than
unity~\cite{Xic-personal-comm}.  In estimating the systematic error
due to this source, the fraction of $\lamll$ candidates due to $\Xib$
decays is varied from $10\%$ to $50\%$.
The $\Lb$ lifetime will also be quoted
with a functional dependence on the level of $\Xib$ contamination in
the $\lamll$ sample, $f_{\Xib/\lamll}$, using the average $\Xib$ lifetime of
$1.39^{+0.34}_{-0.28}\ps$ as measured from $\Xi\ell$
correlations~\cite{Xi_b}.

\subsubsection {Other backgrounds}
 
The $\Ds$ ($\Lc$) candidate may be a misidentified charm hadron if
one or more tracks are assigned the wrong particle type. This
constitutes an additional source of background which is studied using
simulated events. For this background, it is found that the invariant
mass distribution around the $\Ds$ ($\Lc$) mass is similar to that
of the combinatorial background. Such events are therefore considered
to be included in the combinatorial background fraction.

The level of background from genuine $\Ds$ ($\Lc$) particles which
are combined with a hadron that is misidentified as a lepton can be
estimated by fitting the invariant mass spectrum of wrong-sign
combinations in which the charm candidate and the lepton candidate
have the same charge.  This assumes that random
combinations are equally likely to have right and wrong charge
correlations. For each channel where the charm hadron is fully
reconstructed, the wrong sign signal is consistent with zero.
This is in agreement with what has been found in a related analysis that
has greater statistical significance~\cite{opalBzeroBplus}.
This background source is therefore neglected.
 
The background from random associations of a $\Ds$ ($\Lc$) with
genuine leptons is estimated using simulated data. The contribution is 
less than one event to each of our samples
and is neglected.

In the two modes in which the charm hadron is partially reconstructed
from a semileptonic decay channel, $\Ds\to\phi\lp\nu\X$ and
$\Lc\to\Lambda\lm\bar{\nu}\X$, there are additional backgrounds to
consider.  These include the accidental combination of a $\phi$
($\Lambda$), generally produced in fragmentation, with two leptons
which arise from a semileptonic bottom hadron decay, followed by a
semileptonic charm hadron decay.  In the case of the $\Lcl$ channel,
this background is estimated using the observed peak in the $\Lambda$
invariant mass spectrum for the wrong sign combination formed by an
$\lm$ from the bottom hadron decay, an $\lp$ from the charm hadron
decay and a $\overline{\Lambda}$ (whereas signal would be a
$\Lambda$).  Fitting the $\Lambda$ invariant mass distribution for
these wrong sign, $[{\overline\Lambda}\lp]\lm$, candidates, yields a
contribution from this source of $9\pm5$ events.  The contribution
from random associations of a particle that is not from $\Lc$ decay
with a $\Lambda$ and a lepton from $\Lb$ decay is found to be
negligible.  For the analogous $\Ds$ decay into $\phi\lm\overline{\nu}
\mathrm{X}$, there is no wrong sign distribution available, and hence
simulated events are used to estimate this contribution to be
$2.5\pm0.5$ candidates.  The potential contribution of leptons from
$\rm J/\psi$ decays, which are then combined with a $\phi$ (either
from fragmentation or from a b~hadron decay) is also estimated using
simulated events to be $0.5\pm0.3$ candidates. Finally, the
contribution from hadrons misidentified as leptons is estimated by
selecting events in which the two leptons in these modes have the same
sign and is found to contribute $2\pm2$ candidates.
 
The non-combinatorial background sources mentioned above are
expected to contribute a total of $\nbsphysBG$ events to the $\Dsl$ signal 
and  $\nlbphysBG$ events to the $\Lcl$ signal.  
The background subtracted number of $\Dsl$ signal candidates is 
therefore 
$$
  N({\Bs\to\Dslnux})\ = \ \ncandbs \pm \ncandbserr\ . 
$$
The background subtracted number of $\Lcl$ signal candidates is
$$
  N({\Lb\to\Lclnux})\ = \ \ncandlb \pm \ncandlberr\ ,
$$
where no correction has been made for possible $\Xib$ contamination.

\section{The $\bsym \Bs$ and $\bsym \Lb$ Lifetime Fit}

To extract the $\Bs$ ($\Lb$) lifetimes from the measured decay
lengths, an unbinned maximum likelihood fit is performed using a likelihood
function that accounts for both the signal and background components
of the sample.  This fit is largely the same as has been used
previously in similar OPAL measurements~\cite{opalBslife,opal-Lc-Lblife}.


\subsection{Boost Determination}
 
For the component of the likelihood function describing the $\Bs$
($\Lb$) signals, the $\Bs$ ($\Lb$) lifetime must be related to the
observed decay lengths.  Since neither channel is fully reconstructed,
because at least the neutrino produced in the b~hadron decay is not
reconstructed, it is necessary to estimate the b~hadron momentum,
$\pb$.  The probability distribution of a given candidate having a
particular $\Bs$ ($\Lb$) momentum, ${\cal B}$, is estimated on an
event-by-event basis in one of two ways, depending on the decay
channel.

For the semileptonic $\Ds$ decay mode and both $\Lc$ decay modes, the
technique employed in reference~\cite{opal-Lc-Lblife} is used.  This relies on
information from Monte Carlo simulation to estimate the probability
distribution of b~hadron energy, given the observed momentum, $\pdi$,
and invariant mass, $\mdi$ of all the observed tracks in the candidate
(i.e., $\KKll$, $\pKpl$ or $\ppill$).  Using a conversion factor,
$R\equiv\pdi/\pb$, the relationship of the decay time, $t$, to the
decay length $L$, can be expressed as $t = L \cdot R \cdot
({\mb}/{\pdi})$, where $\mb$ is the $\Bs$ ($\Lb$) mass.  The
distribution of $R$ was determined using simulated data for the
signal $\Lc$ decay modes and the semileptonic $\Ds$ decay mode
produced with the JETSET~7.4 Monte
Carlo~\cite{jetset}, using the fragmentation function of Peterson
\etal~\cite{peterson}.  For each decay mode, twelve $R$ 
distributions were produced covering
different ranges of the momentum and invariant mass of the $\Dsl$
($\Lcl$) combination.  These distributions are used in the lifetime
fit to describe the probability, ${\cal B}(\pb \mid \pdi, \mdi)$ that
a candidate with a measured $\pdi$ and $\mdi$ will have a particular
$\Bs$ ($\Lb$) momentum.

For the other $\Ds$ decay modes in which the $\Ds$ is fully
reconstructed, an analytic approach is
used~\cite{opalBslife,opalBzeroBplus}.  This approach is not
applicable in the case of the $\Lb$ decays because of the possibility
of large polarization effects.  The same momentum and invariant mass
observables ($\pdi$ and $\mdi$) as the other method are employed, but
an exact calculation is used, based on the kinematics of the
$\Bs\to\Dslnux$ decay rather than Monte Carlo simulation, to calculate
the probability distribution ${\cal B}(\pb \mid \pdi, \mdi)$.  This
approach is described in detail in reference~\cite{opalBzeroBplus}.

\subsection{Likelihood Functional Form}
 
 
The likelihood function for observing a particular decay length of a
$\Bs$ ($\Lb$) hadron may now be parametrized in terms of the
measurement error of the decay length, the $\Dsl$ ($\Lcl$) invariant
mass and momentum, and the assumed lifetime.  The functional form of
the likelihood is given by the convolution of three terms: an
exponential whose mean is the $\Bs$ ($\Lb$) lifetime, the boost
distribution obtained from the values of the observed $\Dsl$ ($\Lcl$)
mass and momentum, and a Gaussian resolution function with width equal
to the measured decay length error.  This can be expressed as:
\begin{equation}
  \LB(\li \mid \tb ,\sigLi, \pdi, \mdi) =
   \int_0^\infty {\mathrm d}l 
   \int_0^{\pb^{\rm{max}}}{\mathrm d}\pb \ \   
    {\cal G}(\li \mid l,\sigLi) \ \,
    {\cal B}(\pb \mid \pdi, \mdi) \ \,
    {\cal P}(l \mid \tb, \pb) \ ,
\end{equation}
where $\pb^{\rm max}$ is the maximum possible energy that the b hadron can have. 
The function ${\cal G}$ is a Gaussian
function that describes the probability to observe a decay length,
$\li$, given a true decay length $l$ and the measurement uncertainty
$\sigLi$.  ${\cal B}$ is the probability of a particular $\Bs$ ($\Lb$)
momentum for an observed momentum, $\pdi$ and invariant mass, $\mdi$ of
all tracks comprising the candidate.  ${\cal P}$ is the
probability for a given $\Bs$ ($\Lb$) to decay at a distance
$l$ from the $\rm{e}^+\rm{e}^-$ interaction point.  This function is given by:
\begin{equation}
  {\cal P}(l \mid \tb, \pb) = 
    \frac{\mb}{\tb\pb}
        \exp \left[\frac{-l\cdot \mb}{\tb\pb}\right] \ ,
\end{equation}
where $\tb\pb/\mb$ is the mean decay length for a given momentum, 
$\pb$, mean lifetime, $\tb$, and mass, $\mb$, of the $\Bs$ ($\Lb$).  

As discussed previously, non-combinatorial (physics) backgrounds
result from the decay of other b~hadrons.  The likelihood function
describing these sources of background is therefore taken to have the
same form as the $\Bs$ ($\Lb$) signal, except that the b~hadron
lifetimes contributing to these background samples are fixed to the
exclusive world average values~\cite{PDG}, weighted appropriately.
The level of the contributions to this background are set to fixed
fractions of the signal, as determined in the previous section.  The
effects of the uncertainty in these fractions on the lifetime are
addressed as a systematic error.  The likelihood function, $\LDl$,
describing all sources of $\Dsl$ ($\Lcl$) combinations --- the $\Bs$
($\Lb$) signal as well as these physics backgrounds --- is just a
linear combination of $\LB$ and the physics background contributions.
For the semileptonic channels, the lifetime distribution of the
backgrounds which include a real $\phi$ ($\Lambda$) not from a $\Ds$
($\Lc$), is estimated from the sideband, in the case of $\Ds$, or the
sideband and the wrong-sign distribution for the $\Lc$ mode.  For the
purposes of determining the lifetime properties of the background,
these background sources are treated as combinatorial.

 
The fit must also account for the combinatorial background present in
the $\Dsl$ ($\Lcl$) sample. The functional form used to parametrize
this source of background is composed of a positive and a negative
exponential, each convolved with the same boost function and Gaussian
resolution function as the signal.  This can be expressed as,
\begin{center}
$\Lcombi(\li \mid \tbgp,\tbgn,\fp,\sigLi,\pdi,\mdi)
=\hspace{8cm}$
\end{center} 
\begin{equation}
\int_0^\infty {\mathrm d}l 
   \int_0^{\pb^{\rm max}}{\mathrm d}\pb \ \  
    {\cal G}(\li\mid l,\sigLi) \ \,
    {\cal B}(\pb \mid \pdi, \mdi) \ \,
    {\cal P}_{bg}(L \mid \tbgp, \tbgn, \fp, \pb)\ ,
\end{equation}
where 
\begin{equation}
   {\cal P}_{bg}(l \mid \tbgp,\tbgn,\fp, \pb) =
   \fp\,\frac{\mb}{\tbgp\pb}\,
        \exp\left[\frac{-\,l \cdot \mb}{\tbgp\pb}\right]
 + (1-\fp)\,\frac{\mb}{|\tbgn|\pb}\,
      \exp\left[\frac{-\,(-l) \cdot \mb}{|\tbgn|\pb}\right]\  .
\end{equation}
The fraction of background with positive lifetime, $\fp$, as well as
the characteristic positive and negative lifetimes of the background,
$\tbgp$ and $\tbgn$, are free parameters in the fit.  This
double-exponential shape is motivated by considerations of event
topologies that can lead to apparent negative decay lengths, even
before resolution effects are considered.  The background parameters
are fitted separately for the hadronic and semileptonic $\Ds$ ($\Lc$)
modes, since the lifetime properties of the real $\phi$ ($\Lambda$)
backgrounds may well be different from the purely combinatorial
background in the other decay modes.

The background in the event sample is taken into account by
simultaneously fitting for the signal and background contributions.
The probability that a candidate arises from combinatorial background,
$\fbgi$, is determined as a function of the observed invariant mass of
this candidate from the fits to the invariant mass spectra shown in
figures~\ref{fig:Dsmass} and \ref{fig:lcmass}.


Thus, the full likelihood for candidate $i$ is:
\begin{equation}
 {\cal L}_i(\li \mid \tb ,\sigLi, \pdi, \mdi) \ = \
  (1 - \fbgi)\cdot\LDl \  + \  \fbgi\cdot\Lcombi \ .
\end{equation}
In total, four parameters are free in the fit: the $\Bs$ ($\Lb$)
lifetime, and the parameters describing the combinatorial background
($\fp$, $\tbgp$ and $\tbgn$).

\subsection{Lifetime Fit Results}
 

The fit to the decay lengths of the $\ncandDLbs$ $\Dsl$ combinations
yields $\tauBs = \fittaubsRAW \ps$, where the error is statistical
only.  The fit to the $\ncandDLlb$ $\Lcl$ candidates yields $\tauLb =
\fittaulbRAW \ps$.\footnote{In comparing these two measurements, note
  that the fractional errors do not scale directly with
the number of events in the signal because of the larger combinatorial
background in the $\pKpi\ell^-$ mode, which is the statistically dominant 
mode for the $\Lb$ lifetime measurement.}
The results of these fits are shown in
figures~\ref{fig:bslifetimefit} and \ref{fig:lblifetimefit}.  In each
figure, the candidates are divided into two categories --- signal
region and sideband region --- in order to show the behaviour of the
fit when the candidate sample consists mostly of signal or mostly of
combinatorial background.  The signal region is defined as the mass
region within two standard deviations of the fitted invariant mass
(see figures~\ref{fig:Dsmass} and ~\ref{fig:lcmass}).  The curves in
these figures represent the sums of the decay length probability
distributions for each event.  The figures indicate that the fitted
functional forms provide a good description of the data for both
signal and background.  For the $\Bs$ ($\Lb$) fit, a total $\chi^2$ of
5.2 (9.6) is found for the sum of the signal and sideband decay length
distributions for 12 (11) bins that contain at least five candidates.
As was stated earlier, the fits are to unbinned data.
\begin{figure}[ptb]
  \begin{center}
  \begin{minipage}{0.9\textwidth}
    \epsfxsize=\textwidth 
    \epsffile{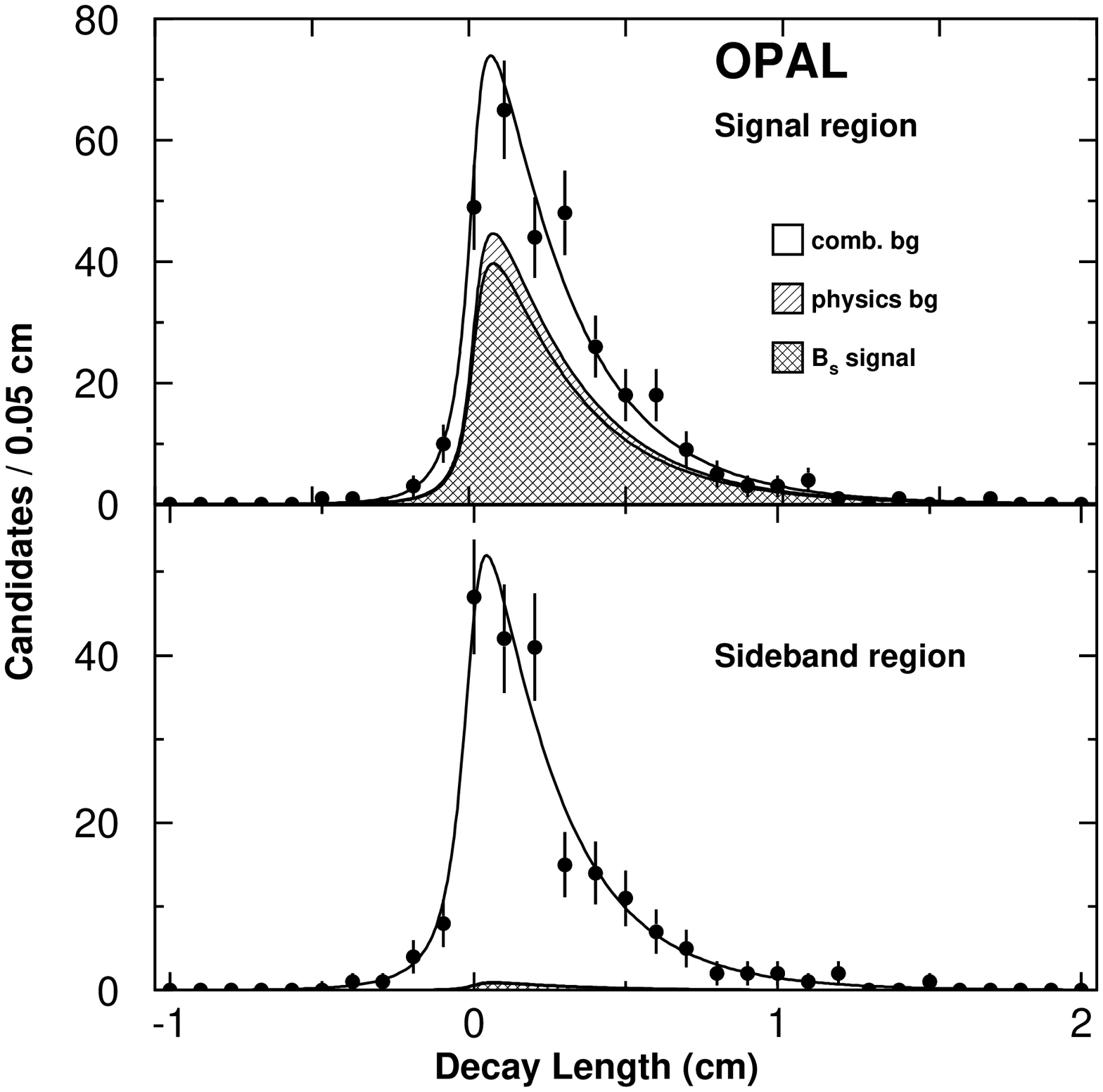}
  \end{minipage}
  \vspace{-0.in} \Caption{Top: The decay length distribution of $\Dsl$
    combinations with an invariant mass within the signal region.  The
    unhatched area represents the contribution from combinatorial
    background, the hatched area is the contribution from sources of
    non-combinatorial background and the double-hatched region is due
    to signal from decays of a $\Bs$.  Bottom: The similar decay
    length distribution for candidates with an invariant mass in the
    sideband region. The curves are the results of the fit described
    in the text. }
     \label{fig:bslifetimefit}
  \end{center}
\end{figure}
\begin{figure}[ptb]
  \begin{center}
  \begin{minipage}{0.9\textwidth}
    \epsfxsize=\textwidth 
    \epsffile{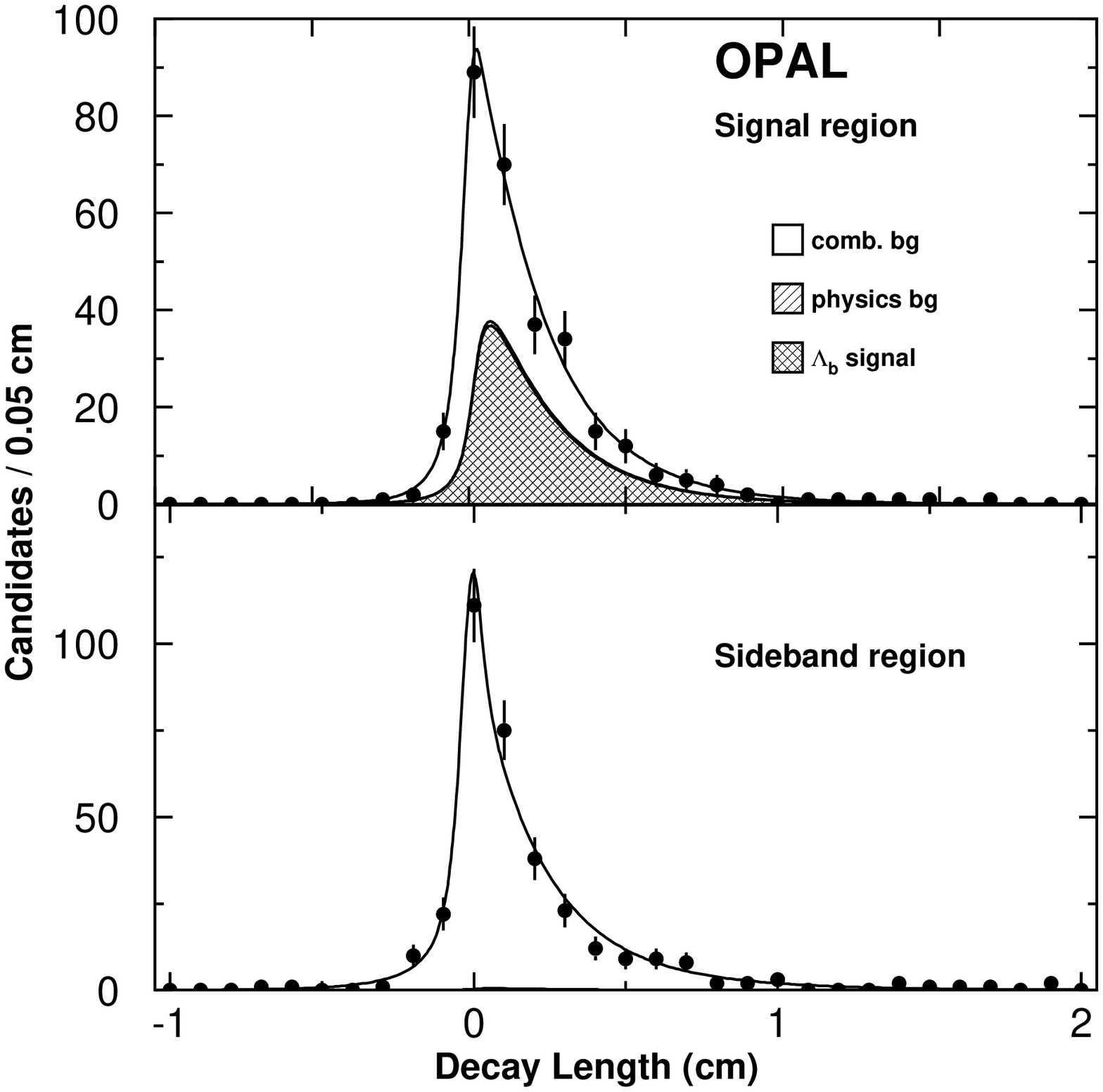}
  \end{minipage}
      \vspace{-0.in} 
      \Caption{Top: The decay length distribution of
        $\Lcl$ combinations with an invariant mass within the signal
        region.  The unhatched area represents the contribution from
        combinatorial background, the (very small) hatched region 
        represents non-combinatorial background, not including
        any $\Xib$ contribution, and the
        double-hatched area is due to signal from decays of a
        $\Lb$. 
        Bottom: The similar decay length distribution for
        candidates with an invariant mass in the sideband region.  The
        curves are the results of the fit described in the text.  }
     \label{fig:lblifetimefit}
  \end{center}
\end{figure}

\section{\label{sec-check} Checks of the Method}



Tests are performed on several samples of simulated events to check for
biases in the selection and fitting procedures.  The first tests
involve a simple Monte Carlo program which generates decay length data
for the signal $\Dsl$ ($\Lcl$) decays and combinatorial background.
For each signal candidate from a $\Bs$ ($\Lb$) decay, this simulation
generates a $\Bs$ ($\Lb$) decay time from an exponential distribution
with the mean set to a known value.  The $\Bs$ ($\Lb$) momenta are
chosen from a spectrum based on the full Monte Carlo simulation.  The
$\Bs$ ($\Lb$) decay length is then calculated and combined with the
momentum to give
the true candidate decay time. This is then degraded by a resolution
function. Physics backgrounds are generated through a similar
procedure.  Many fits are conducted over wide ranges of $\Bs$ ($\Lb$)
lifetimes with different levels and parametrizations of the
backgrounds.  The results of these studies show no biases in the
fitted $\Bs$ ($\Lb$) lifetime to a level of less than 0.5\% and that
the statistical precision of the fit to data is consistent with the
sample size and composition.

To verify that the $\Dsl$ ($\Lcl$) selection does not bias the
reconstructed sample, lifetime measurements are made using simulated
event samples in which large numbers of the decays of interest are
produced.  In these tests it is found that the mean lifetime of the
selected sample of candidates is consistent with the lifetime used to
generate the sample, indicating no bias in the selection procedure.
Applying the lifetime fit to the selected samples similarly shows no
evidence for a bias to within the statistical precision allowed by
these samples.  This precision ranged from 2\% for the $\KKp$ and
$\pKpi$ channels to 4\% for the other decay modes.  Similarly, applying
the selection and fitting procedure to a Monte Carlo simulation sample of 4
million hadronic $\Zzero$ decays, the fitted $\Bs$ ($\Lb$)
lifetimes agree with the generated lifetimes to within the statistical 
power of the sample.

The lifetime fit is also applied to each decay channel
individually. The resulting lifetimes of these separate fits are 
consistent with each other and with the fit to the entire sample.

\section{\label{sec-sys} Evaluation of Systematic Errors}
 
The sources of systematic error considered are those due to the level,
parametrization and source of the background, the boost estimation
method, possible polarization of b baryons, the beam spot determination 
and possible tracking errors.
These systematic errors are summarized in table~\ref{tab-syserr}.
 


The uncertainty in the level of combinatorial background has two
sources: the uncertainty due to the mass fit to the candidate
invariant mass spectra and the statistical variation of the background
under the invariant mass peak.  Background fractions due to one
standard deviation variations for these two cases are determined and
used in the lifetime fit. This produces variations in the $\Bs$ and
$\Lb$ lifetimes of $\pm 0.03\ps$.  The width of the sideband region of
the mass spectra, from which candidates are selected for use in the
lifetime fit, is also varied yielding contributions to the systematic
errors of $\pm 0.01\ps$ for the $\Bs$ and $\pm 0.03\ps$ for the $\Lb$
lifetimes.  Using a quadratic function to describe the mass distribution
of the combinatorial background has a negligible effect on the fitted
lifetimes.

The effect of the uncertainty in the level of the non-combinatorial
backgrounds to the $\Bs$ and $\Lb$ signal is estimated by varying
these background levels over the ranges described in
section~\ref{sec:phys-bg}. This produces a $\pm0.01\ps$ variation in
the $\Bs$ lifetime. The systematic shift due to a $\Xib$ contamination
of $30\pm20\%$ in the $\Lb$ sample is $\XibLbsys\ps$.  The shift in
the central value of the measured lifetime as a function of $\Xib$
contamination is $+0.08\cdot f_{\Xib/\lamll}\ps$.  All other
non-combinatorial backgrounds mentioned in section~\ref{sec:phys-bg}
contribute an additional error of $\pm0.01\ps$ to the $\Lb$ lifetime.
The b~hadron lifetimes used for these backgrounds are also varied
within their measured errors~\cite{PDG}.  The resulting change in the
fitted lifetimes is less than $\pm0.01\ps$.

The total systematic error associated with the description of the
combinatorial and physics backgrounds is therefore $\pm0.03\ps$
($\pm0.05\ps$) for the measured $\Bs$ ($\Lb$) lifetimes.

The effects of uncertainty in the b~hadron fragmentation are estimated
slightly differently for the two boost estimation methods employed.
For the hadronic $\Ds$ decay modes, the estimated $\Bs$ energy
spectrum used by the boost estimation procedure is varied within the
measured limits of the average b~hadron energy~\cite{lep-avg-x_b}.
Similarly, in generating the Monte Carlo events used to estimate the
b~hadron boost for the two $\Lc$ decay modes and the $\phil$ mode,
the average b~hadron energy was varied by the same amounts as used  above.
These variations yield changes in the fitted lifetimes of $\pm0.02\ps$.
The effect on the lifetime of varying the mass of the $\Lb$ by
$\pm50\,\MeVcc$ about its central value of $5641\,\MeVcc$~\cite{PDG}
is $\pm 0.01\ps$.  The effect of a $2\,\MeVcc$ uncertainty in the mass
of the $\Bs$~\cite{PDG} results in a change of less than $0.01\ps$ in
the measured $\Bs$ lifetime.

In the Monte Carlo events used for the boost estimate, the
$\Lb$ was assumed to be unpolarized.  However, in the Standard Model,
b baryons can retain up to the full longitudinal polarization of
$-0.94$ from the b quark.  A variation from 0 to $-0.94$ polarization
produces a change of $+0.06\ps$.  A recent measurement of the
polarization~\cite{lbpol} is used to correct the lifetime extracted
using the decay length fit.  This yields a correction of
$+0.014^{+0.020}_{-0.014}\ps$, where this error results from the
precision of the polarization measurement.  The effect of the choice
of form factor used to describe the energy transfer from the $\Lb$ to
the $\Lc$ has also been investigated.  The use of the alternative form
factors of reference~\cite{formf} produces a negligible change in the
fitted lifetime.

The average interaction point of the LEP beams in OPAL is used as the
estimate of the production vertex of the $\Bs$ and $\Lb$ candidates.
The mean coordinates of the beam spot are known to better than
$25\mic$ in the $x$ direction and $10\mic$ in $y$.  The effective
\mbox{r.m.s.} spread of the beam is known to a precision of better
than $10\mic$ in both directions.  To test the sensitivity of \tauBs\ 
and \tauLb\ to the assumed position and size of the beam spot, the
coordinates of the beam spot are shifted by $\pm 25\mic$, and the
spreads are changed by $\pm 10\mic$.  The largest observed variation
in \tauBs\ and \tauLb\ is $0.01\ps$ which is assigned as a systematic
error to both measurements.
 
The effects of alignment and calibration uncertainties on the result
are not studied directly but are estimated from a detailed study of
3-prong $\mathrm{\tau}$ decays~\cite{taulife}, in which the uncertainty
in the decay length due to these effects is found to be less than
$1.8\%$ for the data taken during 1990 and 1991 and less than $0.4\%$
for later data. This corresponds to an uncertainty on \tauBs\ and
\tauLb\ of $0.01\ps$.  The potential for incorrect estimation of the decay
length error is addressed by allowing an additional parameter in
the lifetime fit which is a scale factor on the estimated  decay length
error.  This parameter is found to be consistent with unity.
This procedure changes the $\Bs$ lifetime by less than $0.01\ps$ and
the $\Lb$ lifetime by $-0.02\ps$.
 
\begin{table}[thb]
  \centering
  \begin{tabular}{|l|r|r|}
    \hline
      Source  &  $\tauBs$ correction (ps) & $\tauLb$ correction (ps)\\
    \hline
      background (excl. $\Xib$) & 0.00 $\pm 0.03$  &  0.00 $\pm 0.05$ \\
      $\Xib$ background       &                  &  $\XibLbsys$     \\
      uncertainty in boost    & 0.00 $\pm 0.02$  &  0.00 $\pm 0.02$\\
      polarization            &                  & +0.01 $\pm 0.02$ \\
      beam spot               & 0.00 $\pm 0.01$  &  0.00 $\pm 0.01$\\
      alignment errors        & 0.00 $\pm 0.01$  &  0.00 $\pm 0.01$\\
    \hline
      total              &  $\pm \tauBssys$      & +0.03 $\pm \tauLbsys$\\
    \hline
  \end{tabular}
  \Caption{Summary of systematic corrections and uncertainties 
           on the $\Bs$ and $\Lb$ lifetimes.}
  \label{tab-syserr}
\end{table}
 
\section{Conclusion}
 
The decay channels $\Bs\to\Dslnux$ and $\Lb\to\Lclnux$, where $\Ds$
decays to $\KstK$, $\phipi$, $\KKo$ or $\philnux$ and $\Lc$ decays to
$\pKpi$ or $\lamlnux$ have been reconstructed.  From almost
\nGPMH~million hadronic $\Zzero$ events recorded by OPAL from 1990 to
1995, a total of $\ncandbs \pm \ncandbserr$ such candidates are
attributed to $\Bs$ decays and $\ncandlb \pm \ncandlberr$ such
candidates are attributed to $\Lb$ decays.

The $\Bs$ lifetime is found to be 
\begin{center} 
  $\tauBs = \fittaubs \pm \tauBssys  \ps , $
\end{center}
where the first error is statistical and the second systematic.  As
predicted by theoretical calculations~\cite{BIGI,NEUBERT}, this result
is consistent with the observed value for the $\Bd$
lifetime~\cite{PDG}.  This is also in agreement with other
measurements of the $\Bs$
lifetime~\cite{opalBslife-DsInclusive,bslifetime-no-opal}.  The above
value of $\tauBs$ has been combined with the OPAL measurement of the
$\Bs$ lifetime in which only a $\Ds$ candidate is
reconstructed~\cite{opalBslife-DsInclusive} which yields $\tauBs =
1.72^{+0.20}_{-0.19}(\mathrm{stat})^{+0.18}_{-0.17}(\mathrm{syst})\ps$.
Taking the correlated statistical and systematic errors into account,
the average of these two measurements is found to be $1.57\pm0.14\ps$.

The measured $\Lb$ lifetime, is
\begin{center} 
  $\tauLb = (\fittauLbXibval + 0.08\cdot f_{\Xib/\lamll})
   \fittauLbXierr  \pm \tauLbsys \ps$\ , 
\end{center}
where the dependence of the fitted $\Lb$ lifetime is given in terms of
the fraction, $f_{\Xib/\lamll}$, of the $\lamll$ candidates that are
due to $\Xib$ decays, the first error is statistical and the second
systematic.  Assuming $\Xib$ contamination of $30\pm20\%$, the $\Lb$
lifetime is 
\begin{center}
$\tauLb = \fittauLbval\fittauLberr \pm \tauLbsys \ps$. 
\end{center}
The lifetime
using the $\lamll$ sample alone is found to be
$\fittauLblaml\tauLblamllsys\ps$, when the $\Xib$ content is taken 
to be $30\pm20\%$. For the $\pKpi\ell^-$ sample alone, the $\Xib$ content
is estimated to be only about 1\% and the lifetime of this sample is
$\fittauLbpKpi\pm\tauLbsys\ps$.  These are consistent with other
recent
measurements~\cite{opal-Lamlepton-Lblife,lambda_b-lifetime-no-opal}
which have tended to be lower than the $\Bd$ meson lifetime in
qualitative agreement with the predictions of~\cite{BIGI,NEUBERT}.

\par
\section*{Acknowledgements:}
\par
We particularly wish to thank the SL Division for the efficient operation
of the LEP accelerator at all energies
 and for
their continuing close cooperation with
our experimental group.  
We also thank I.I.\ts Bigi, B.\ts Melic, M.\ts Neubert and M.B.\ts Voloshin
for their aid in understanding the present
theoretical description of charmed baryon decays.
We thank our colleagues from CEA, DAPNIA/SPP,
CE-Saclay for their efforts over the years on the time-of-flight and trigger
systems which we continue to use.  In addition to the support staff at our own
institutions we are pleased to acknowledge the  \\
Department of Energy, USA, \\
National Science Foundation, USA, \\
Particle Physics and Astronomy Research Council, UK, \\
Natural Sciences and Engineering Research Council, Canada, \\
Israel Science Foundation, administered by the Israel
Academy of Science and Humanities, \\
Minerva Gesellschaft, \\
Benoziyo Center for High Energy Physics,\\
Japanese Ministry of Education, Science and Culture (the
Monbusho) and a grant under the Monbusho International
Science Research Program,\\
German Israeli Bi-national Science Foundation (GIF), \\
Bundesministerium f\"ur Bildung, Wissenschaft,
Forschung und Technologie, Germany, \\
National Research Council of Canada, \\
Research Corporation, USA, \\
Hungarian Foundation for Scientific Research, OTKA T-016660, 
T023793 and OTKA F-023259.



\begin{thebibliography}{10}

\bibitem{PDG}
{ Particle Data Group, R.M.~Barnett et al., Phys.~Rev. {\bf D 54} (1996)~1.}

\bibitem{HQET}
{ G.\ts Altarelli and S.\ts Petrarca, Phys. Lett. {\bf B 261} (1991) 303;\\
  I.\ts Bigi, Phys. Lett. {\bf B 169} (1986) 101;\\ J.H.\ts K\"uhn et al., {\sl
  Heavy Flavours at LEP}, MPI--PAE/PTh 49/89, August 1989, contribution by R.
  R\"uckl, p.~59.}

\bibitem{BIGI}
{ I.\ts Bigi, Nuovo Cim. {\bf 109A} (1996) 713;\\ I.\ts Bigi et al.,
  (CERN-TH.7132/94), from the second edition of the book `B Decays,' S.\ts
  Stone (ed.), World Scientific, pp.\ts~132-157;\\ I.I.\ts Bigi and N.G.\ts
  Uraltsev, Phys. Lett. {\bf B 280} (1992) 271.}

\bibitem{NEUBERT}
{ M.\ts Neubert and C.T.\ts Sachrajda, Nucl. Phys. {\bf B 483} (1997) 339.}

\bibitem{opalBslife}
{ OPAL Collab., R.\ts Akers et al., Phys. Lett. {\bf B 350} (1995) 273 .}

\bibitem{opalBslife-DsInclusive}
{ OPAL Collab., K.\ts Ackerstaff et al., ``A Measurement of the $\rm B^0_s$
  Lifetime using Reconstructed $\rm D_s^-$ Mesons'', CERN-PPE/97-095, submitted
  to Z. Phys. C.}

\bibitem{bslifetime-no-opal}
{ ALEPH Collab., D. Buskulic et al., Phys. Lett. {\bf B 377} (1996) 205;\\
  ALEPH Collab., D. Buskulic et al., Z. Phys, {\bf C 69} (1996) 585;\\ CDF
  Collab., F. Abe et al., Phys. Rev. Lett. {\bf 77} (1996) 1945;\\ CDF Collab.,
  F. Abe et al., Phys. Rev. Lett. {\bf 74} (1995) 4988;\\ DELPHI Collab., P.
  Abreu et al., Z. Phys. {\bf C 71} (1996) 11.}

\bibitem{opal-Lc-Lblife}
{ OPAL Collab., R.\ts Akers et al., Phys Lett. {\bf B 353} (1995) 402.}

\bibitem{opal-Lamlepton-Lblife}
{ OPAL Collab., R.\ts Akers et al., Z. Phys. {\bf C 69} (1996) 195.}

\bibitem{lambda_b-lifetime-no-opal}
{ ALEPH Collab., D. Buskulic et al., ``Measurement of the b baryon lifetime and
  branching fractions in Z decays'', CERN-PPE/97-111, submitted to Z. Phys.
  C;\\ CDF Collab., F. Abe et al., Phys. Rev. Lett. {\bf 77} (1996) 1439;\\
  DELPHI Collab., P. Abreu et al., Z. Phys. {\bf C 71} (1996) 199;\\ DELPHI
  Collab., P. Abreu et al., Z. Phys. {\bf C 68} (1995) 375.}

\bibitem{opaldet-opalsi-opalsi2}
{ OPAL Collab., K.\ts Ahmet et al., Nucl. Inst. and Meth. {\bf A 305} (1991)
  275;\\ P.P.\ts Allport et al., Nucl. Inst. and Meth. {\bf A 324} (1993) 34;\\
  P.P.\ts Allport et al., Nucl. Inst. and Meth. {\bf A 346} (1994) 476.}

\bibitem{opaljet-hid}
{ O.\ts Biebel et al., Nucl. Inst. and Meth. {\bf A 323} (1992) 169;\\ M.\ts
  Hauschild et al., Nucl. Inst. and Meth. {\bf A 314} (1992) 74.}

\bibitem{jetset}
{ T. Sj\"{o}strand, Comp. Phys. Comm. {\bf 82} (1994) 74. \\ The OPAL parameter
  optimization is described in \\ OPAL Collab., G.\ts Alexander et al., Z.
  Phys. {\bf C 69} (1996) 543.}

\bibitem{peterson}
{ C. Peterson et al., Phys. Rev. {\bf D 27} (1983) 105.}

\bibitem{opalmc}
{ J.\ts Allison et al., Nucl. Inst. and Meth. {\bf A 317} (1992) 47.}

\bibitem{jetpol}
{ A special subroutine for the decays of b-flavoured baryons was provided by T.
  Sj\"{o}strand.}

\bibitem{formf}
{ X.H.\ts Guo and P.\ts Kroll, Z. Phys. {\bf C 59} (1993) 567.}

\bibitem{opalmh}
{ OPAL Collab., G.\ts Alexander et al., Z. Phys. {\bf C 52} (1991) 175.}

\bibitem{bib-JADE}
{ JADE Collab., W.\ts Bartel et al., Z.~Phys. {\bf C 33} (1986) 23;\\ JADE
  Collab., S.\ts Bethke et al., Phys. Lett. {\bf B 213 }(1988) 235.}

\bibitem{opal-track-sel}
{ OPAL Collab., R.\ts Akers et al., Phys. Lett. {\bf B 316} (1993) 435.}

\bibitem{BdmixNNeID}
{ OPAL Collab., R.\ts Akers et al., Phys. Lett. {\bf B 327} (1994) 411.}

\bibitem{gambblep}
{ OPAL Collab., P.D.\ts Acton et al., Z. Phys. {\bf C 58} (1993) 523.}

\bibitem{kssel}
{ OPAL Collab., R.\ts Akers et al., Z. Phys. {\bf C 67} (1995) 389.}

\bibitem{taulife}
{ OPAL Collab., P.D.\ts Acton et al., Z. Phys. {\bf C 59} (1993) 183; \\ OPAL
  Collab., R.\ts Akers et al., Phys. Lett. {\bf B 338} (1994) 497.}

\bibitem{cleo-BtoLamc}
{ CLEO Collab., R.\ts Ammar et al., Phys. Rev. {\bf D 55} (1997) 13}.

\bibitem{Xic-theory}
{ B.\ts Guberina and B.\ts Melic, ``Inclusive Charmed-Baryon Decays and
  Lifetimes'', IRB-TH 1/97, April 1997;\\ M.B.\ts Voloshin, Phys. Lett. {\bf B
  385} (1996) 369.}

\bibitem{argus-BtopeX}
{ ARGUS Collab., H.\ts Albrecht et al., Phys. Lett. {\bf B 249} (1990) 359. The
  decay used in this reference to measure the branching ratio $\Br(\Bbar_{\rm
  u,d}\to\p\lm\nubar\X)$ is $\B\to\bar{\p}\mathrm{e}^+\nu\X$.}

\bibitem{Xi_b}
{ ALEPH Collab., D.\ts Buskulic et al., Phys. Lett. {\bf B 384} (1996) 449;\\
  DELPHI Collab., P.\ts Abreu et al., Z. Phys. {\bf C 68} (1995) 541.}

\bibitem{b-baryon-masses}
{W.\ts Kwong and J.L.\ts Rosner, Phys. Rev. {\bf D 44} (1991) 212.}

\bibitem{Xic-personal-comm}
{ Personal communications with I.I.\ts Bigi, B.\ts Melic, M.\ts Neubert and
  M.B.\ts Voloshin.}

\bibitem{opalBzeroBplus}
{ OPAL Collab., R.\ts Akers et al., Z. Phys. {\bf C 67} (1995) 379.}

\bibitem{lep-avg-x_b}
{ The LEP Collaborations, ALEPH, DELPHI, L3 and OPAL, and the LEP Electroweak
  Working Group, Nucl. Inst. and Meth. {\bf A 378} (1996) 101.}

\bibitem{lbpol}
{ ALEPH Collab., D. Buskulic et al., Phys. Lett. {\bf B 365} (1996) 437.}

\end{thebibliography}
\end{document}